\documentstyle[prl,aps,epsf,twocolumn]{revtex}
\def\urs{URu$_2$Si$_2$}
\def\cecu6{CeCu$_6$}
\def\upt3{UPt$_3$}
\def\mbohr{$\mu_{\rm B}$}
\def\mord{$\mu_{\rm o}$}
\def\pcr{$P_{\rm c}$}
\def\tord{$T_{\rm o}$}
\def\taf{$T_{\rm m}$}
\def\tafp{$T_{\rm m}^{+}$}
\def\tafm{$T_{\rm m}^{-}$}

\begin{document}
\draft
\wideabs{
\title{
Effect of Pressure on Tiny Antiferromagnetic Moment in the Heavy-Electron 
Compound URu{\boldmath $_{2}$}Si{\boldmath $_{2}$}
}
\author{
H. Amitsuka$^{1}$, M. Sato$^{2}$, N. Metoki$^{2}$, M. Yokoyama$^{1}$, 
K. Kuwahara$^{1}$, T. Sakakibara$^{1}$ \\
H. Morimoto$^{3}$, S. Kawarazaki$^{3}$, 
Y. Miyako$^{3}$, and J.A. Mydosh$^{4}$
}
\address{
$^{1}$Graduate School of Science, Hokkaido University, 
Sapporo 060-0810, Japan\\
$^{2}$Advanced Science Research Center, Japan Atomic Energy Research 
Institute, Ibaraki 319-1195, Japan\\
$^{3}$Graduate School of Science, Osaka University, 
Toyonaka 560-0043, Japan\\
$^{4}$Kamerlingh Onnes Laboratory, Leiden University, PO Box 9504, 
2300 RA Leiden, The Netherlands
}
\date{Received 6 July 1999}
\maketitle
\begin{abstract}
We have performed elastic neutron-scattering experiments on the 
heavy-electron compound {\urs} for pressure $P$ up to 2.8 GPa.
We have found that the antiferrmagnetic (100) Bragg reflection 
below {\taf} $\sim$ 17.5 K is strongly enhanced by applying pressure.
For $P <$ 1.1 GPa, the staggered moment {\mord} at 1.4 K 
increases linearly from $\sim$ 0.017(3) {\mbohr} to $\sim$ 
0.25(2) {\mbohr}, while {\taf} increases slightly at a rate $\sim$ 1 K/GPa, 
roughly following the transition temperature {\tord} determined from 
macroscopic anomalies.
We have also observed a sharp phase transition at {\pcr} 
$\sim$ 1.5 GPa, above which a 3D-Ising type of antiferromagnetic phase 
({\mord} $\sim$ 0.4 {\mbohr}) appears with a slightly reduced 
lattice constant.
\end{abstract}
}
%
%
\narrowtext
Antiferromagnetism due to extremely weak moments 
indicated in {\cecu6}\cite{refcc6}, 
{\upt3}\cite{aeppli88} and {\urs}\cite{broholm87} has 
been one of the most intriguing issues in heavy-fermion physics.
{\urs} has received special attention because of its unique 
feature that the development of the tiny staggered moment 
{\mord} is accompanied by significant 
anomalies in bulk properties\cite{palstra85,schlabitz86,anne86}.
In particular, specific heat shows a large jump 
$({\Delta}C/T_{\rm o} \sim 300 {\rm mJ/K^{2}mol})$ at {\tord} $=$ 17.5 K, 
which evidences 5f electrons to undergo a phase transition
\cite{palstra85,schlabitz86}.
Microscopic studies of neutron 
scattering\cite{broholm87,mason90,walker93} 
and X-ray magnetic scattering\cite{isaacs90} have revealed an ordered 
array of 5f magnetic dipoles along the tetragonal $c$ axis with a wave 
vector $Q = (100)$ developing below {\tord}.
The magnitude of {\mord} is found to be 0.02--0.04 {\mbohr}, 
which however is roughly 50 times smaller than that of the 
fluctuating moment $(\mu_{\rm para} \sim 1.2 \mu_{\rm B})$ 
above {\tord}\cite{mason95}.
This large reduction of the 5f moment below {\tord} is 
apparently unreconciled with the large $C(T)$ anomaly, 
if $\mu_{\rm para}$ is simply regarded as full moment\cite{buyers96}.

To solve the discrepancy, various ideas have been 
proposed, which can be classified into two groups: 
(i) the transition is uniquely caused by magnetic dipoles with highly 
reduced $g$-values\cite{ge87,sikkema96,okuno98}; and 
(ii) there is hidden order of a non-dipolar degree of freedom
\cite{miyako91,bar93,santini94,ami94,ohkawa99,bar95,kasuya97,ikeda98}.
The models of the second group ascribe the tiny moment to side effects, 
such as secondary order, dynamical fluctuations and coincidental order of 
a parasitic phase.
Each of the dipolar states may have its own energy scale, 
and to take account of this possibility we define {\taf} as the onset 
temperature of {\mord}, distinguishing it from {\tord}.

The crux of the problem will be how {\mord} relates to the 
macroscopic anomalies.
Recent high-field studies\cite{mason95,mentink96,niels97,mentink97} 
have suggested that {\tord} and {\mord} are not 
scaled by a unique function of field. 
In addition, the comparison of {\tord} and {\taf} for the same sample has 
suggested that {\taf} becomes lower than {\tord} in the absence of 
annealing\cite{fak96}. 
In this Letter we have studied the influence of pressure on the tiny moment 
of {\urs}, for the first time, by means of elastic neutron scattering. 
Previous measurements of resistivity and 
specific heat in $P$ up to 8 GPa have shown that 
the ordered phase is slightly stabilized by pressure, 
with a rate of d{\tord}/d$P \sim$ 1.3 K/GPa 
\cite{louis86,mcelfresh87,fisher90,ido93,schmidt93,oomi94}. 
We now show that pressure dramatically increases {\mord} 
and causes a new phase transition.

A single-crystalline sample of {\urs} was grown by the Czochralski 
technique in a tri-arc furnace. 
The crystal was shaped in a cylinder along the $c$ axis 
with approximate dimensions 5 mm diameter by 8 mm long, and 
vacuum annealed at 1000 $^{\circ}$C for one week. 
Pressure was applied by means of a barrel-shaped piston cylinder device 
\cite{onodera87} at room temperature, which was then cooled in a 
$^{4}$He cryostat for temperatures between 1.4 K and 300 K.
A solution of Fluorinert 70 and 77 (Sumitomo 3M Co.\ Ltd., Tokyo) of 
equal ratio served as the quasihydrostatic pressure transmitting medium.
The pressure was monitored by measuring the lattice constant of NaCl, 
which was encapsuled together with the sample.

The elastic neutron-scattering experiments were performed on the 
triple-axis spectrometer TAS-1 at the JRR-3M reactor of Japan Atomic 
Energy Research Institute.
Pyrolytic graphite PG(002) crystals were used for monochromating and 
analyzing the neutron beam with a wavelength $\lambda =$ 2.3551 {\AA}.
We used a 40'-80'-40'-80' horizontal collimation, 
and double 4-cm-thick PG filters as well as 
a 4-cm-thick Al$_{2}$O$_{3}$ filter 
to reduce higher-order contamination.
The scans were performed in the ($hk0$) scattering plane, 
particularly on the antiferromagnetic Bragg reflections 
(100) and (210), and on the nuclear ones (200), (020) and (110).
The lattice constant $a$ of our sample at 1.4 K at ambient pressure is 
4.13(1) {\AA}.
\begin{figure}
\epsfysize=7.5cm
\centerline{\epsffile{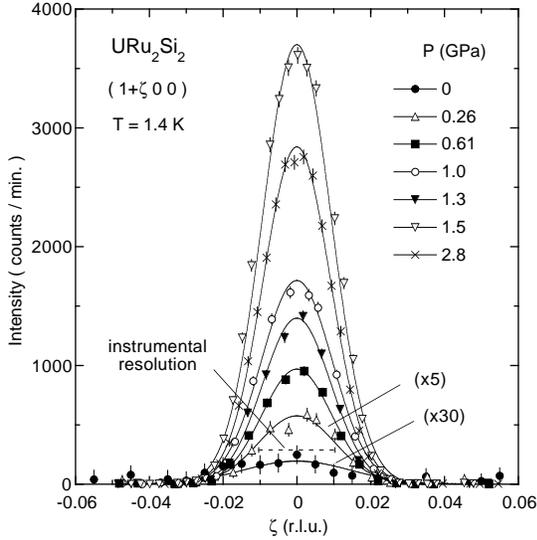}}
\caption{
Longitudinal scans of the antiferromagnetic Bragg peak (100) of {\urs} 
for several pressures.
}
\label{profile}
\end{figure}

Figure~\ref{profile} shows the pressure variations of elastic scans 
at 1.4 K along the $a^{\ast}$ direction through the forbidden nuclear (100) 
Bragg peak.
The instrumental background and the higher-order contributions of nuclear 
reflections were determined by scans at 35 K and subtracted from 
the data.
The (100) reflection develops rapidly as pressure is applied.
No other peaks were found in a survey along the principle axes of the 
first Brillouin zone; 
in addition, the intensities of (100) and (210) reflections follow the $|Q|$ 
dependence expected from the U$^{4+}$ magnetic 
form factor\cite{frazer65} by taking the polarization factor unity.
These ensure that the type-I antiferromagnetic structure at $P = 0$ 
is unchanged by the application of pressure.

The widths (FWHM) of the (100) peaks for $P =$ 0 and 0.26 GPa are 
significantly larger than the instrumental resolution ($\sim$ 0.021(1) reciprocal-lattice 
units), which was determined from $\lambda/2$ reflections at (200).
From the best fit to the data by a Lorentzian function convoluted with 
the Gaussian resolution function, the correlation length 
$\xi$ along the $a^{\ast}$ direction is estimated to be about 
180 {\AA} at $P = 0$ and 280 {\AA} at 0.26 GPa. 
For the higher pressures $P \geq 0.61$ GPa, the simple fits give 
$\xi > 10^{3}$ {\AA}, indicating that the line shapes are resolution limited.

\begin{figure}
\epsfysize=8cm
\centerline{\epsffile{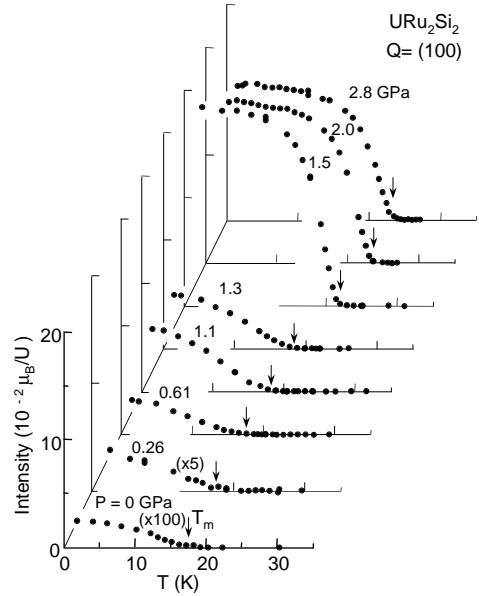}}
\caption{
Temperature dependence of integrated intensity of the (100) magnetic 
Bragg reflection for various pressures.
}
\label{ivst}
\end{figure}

The temperature dependence of the integrated 
intensity $I(T)$ at (100) varies significantly as $P$ traverses 1.5 GPa 
($\equiv$ {\pcr}) (Fig.~\ref{ivst}).
For $P <$ {\pcr}, the onset of $I(T)$ is not sharp: 
$I(T)$ gradually develops at a temperature {\tafp}, which is higher 
than {\tord}, and shows a $T$-linear behavior below a lower temperature 
{\tafm}.
Here we empirically define the ``antiferromagnetic transition" temperature 
{\taf} by the midpoint of {\tafp} and {\tafm}.
The range of the rounding, ${\delta}T_{\rm m} \equiv$ {\tafp} $-$ {\tafm}, is estimated to be 2--3 K, which is too wide to be usual critical scattering. 
Above {\pcr}, on the other hand, the transition becomes sharper 
(${\delta}T_{\rm m} <$ 2 K), accompanied by an abrupt increase in {\taf} 
at {\pcr}. 

\begin{figure}
\epsfysize=8cm
\centerline{\epsffile{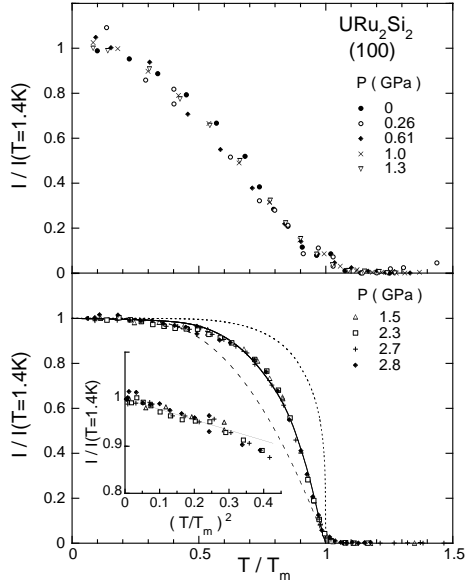}}
\caption{
Normalized intensities $I/I(1.4 {\rm K})$ plotted as a function of 
$T/T_{\rm m}$ for $P <$ {\pcr} (top) and $P >$ {\pcr} (bottom). 
Theoretical calculations based on 2D [39], 3D [35] and mean-field Ising models are also given by dotted, solid and broken lines. 
The inset plots $I/I(1.4 {\rm K})$ versus $(T/T_{\rm m})^{2}$ 
at low temperatures. 
The thin line is a guide to the eye.
}
\label{nivst}
\end{figure}

If $I(T)$ is normalized to its value at 1.4 K, it scales with 
$T/T_{\rm m}$ for various pressures on each side of {\pcr} 
(Fig.~\ref{nivst}). 
This indicates that two homogeneously ordered phases are separated 
by a (probably first-order) phase transition at {\pcr}.
The growth of $I(T)$ for $P <$ {\pcr} is much weaker than that expected for 
the mean-field Ising model, showing an unusually slow saturation of the 
staggered moment. 
On the other hand, the overall feature of $I(T)$ for $P >$ {\pcr} is approximately described by a 3D-Ising model\cite{burley60}. 
In the low temperature range $T/T_{\rm m} < 0.5$, however, 
$I(T)$ rather follows a $T^{2}$ function (the inset of Fig.~\ref{nivst}), 
indicating a presence of gapless collective excitations\cite{kubo52}.

\begin{figure}
\epsfysize=10.5cm
\centerline{\epsffile{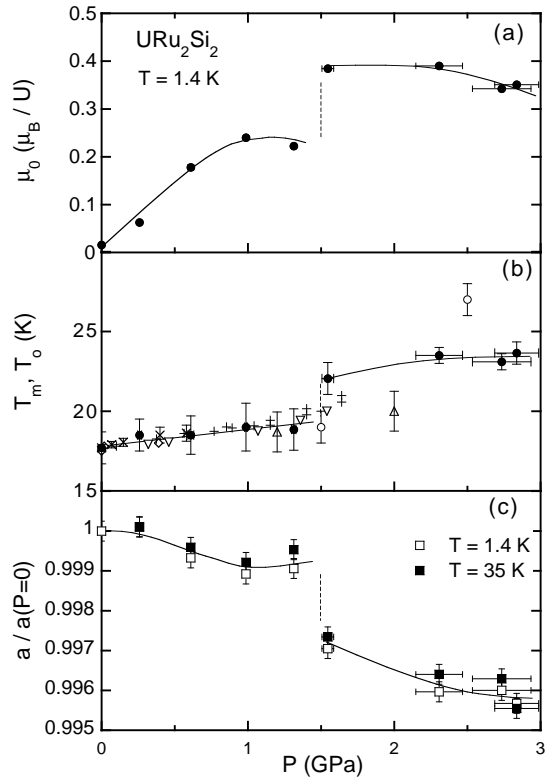}}
\caption{
Pressure variations of (a) staggered magnetic moment {\mord} at 1.4 K; 
(b) the onset temperature {\taf} of the moment determined from 
this work ($\bullet$) and the transition temperature 
{\tord} determined from resistivity 
($\Diamond$ [27], $\bigtriangledown$ [28], 
$\bigtriangleup$ [30], $+$ [31], $\circ$ [32]) 
and specific heat ($\times$ [29]); 
and (c) the relative lattice constant $a(P)/a(0)$ at 1.4 K and 35 K.
{\taf} is defined by $(T_{m}^{+} + T_{m}^{-})/2$ (see the text), and the range 
${\delta}T_{m} (\equiv T_{m}^{+} - T_{m}^{-})$ is shown by using error bars.
The lines are guides to the eye.}
\label{parameters}
\end{figure}

In Fig.~\ref{parameters}, we plot the pressure dependence of {\mord}, 
{\taf} and the lattice constant $a$. 
The magnitude of {\mord} at 1.4 K is obtained through 
the normalization of the integrated intensity at (100) with respect to 
the weak nuclear Bragg peak at (110).
The variation of the (110) intensity with pressure is small ($<$ 5\%) 
and independent of the crystal mosaic, so that the influence of extinction on 
this reference peak is negligible.
{\mord} at $P = 0$ is estimated to be about 0.017(3) {\mbohr}, which is 
slightly smaller than the values ($\sim$ 0.02--0.04 {\mbohr}) of previous 
studies\cite{broholm87,mason90,isaacs90,fak96}, probably 
because of a difference in the selection of reference peaks.
As pressure is applied, {\mord} increases linearly at a rate 
$\sim$ 0.25 {\mbohr}$/$GPa, 
and shows a tendency to saturate at $P \sim$ 1.3 GPa. 
Around {\pcr}, {\mord} abruptly increases from 0.23 {\mbohr} 
to 0.40 {\mbohr}, and then slightly decreases. 

In contrast to the strong variation of {\mord}, 
{\taf} shows a slight increase from 17.7 K to 18.9 K, 
as $P$ is increased from 0 to 1.3 GPa. 
A simple linear fit of {\taf} in this range yields a rate 
$\sim 1.0$ K/GPa, which roughly follows 
the reported $P$-variations of {\tord}. 
Upon further compression, {\taf} jumps to 22 K at {\pcr}, 
showing a spring of $\sim 2.8$ K from a value 
($\sim$ 19.2 K) extrapolated with the above fit. 
For $P >$ {\pcr}, {\taf} again gradually increases 
and reaches $\sim$ 23.5 K around 2.8 GPa.
The pressure dependence of {\tord} in this range is less clear, 
and the few available data points deviate from the behavior of {\taf}, 
see Fig.~\ref{parameters}(b).

The lattice constant $a$, which is determined from the scans at (200), 
decreases slightly under pressure (Fig.~\ref{parameters}(c)). 
From a linear fit of $a$ at 1.4 K for $P <$ {\pcr}, we derive 
$- {\partial}\ln a/{\partial}P \sim 6.7 \times 10^{-4}$ GPa$^{-1}$. 
If the compression is isotropic, this yields 
an isothermal compressibility $\kappa_{\rm T}$ of 
$2 \times 10^{-3}$ GPa$^{-1}$, which is about 4 times smaller 
than what was previously estimated from the compressibilities 
of the constituent elements\cite{anne86}. 
Around {\pcr}, the lattice shrinks with a discontinuous change 
of $- {\Delta}a/a \sim$ 0.2 \%. 
Assuming again the isotropic compression, we evaluate 
${\Delta}\ln V/{\Delta}\mu_{\rm o} \sim - 0.04 \mu_{\rm B}^{-1}$ 
and 
${\Delta}\ln T_{\rm m}/{\Delta}\ln V \sim - 27$ 
associated with this transition. 
Note that a similar lattice anomaly at {\pcr} is observed at 35 K, 
much higher than {\tord}. 
This implies that the system has another energy scale 
characteristic of the volume shrinkage in the paramagnetic region. 
We have confirmed the absence of any lowering 
in the crystal symmetry at {\pcr} within the detectability limit of 
$|a - b|/a \sim 0.05$ \% and 
$\cos^{-1}(\widehat{a}\cdot\widehat{b}) \sim 2'$. 
The $c$ axis is perpendicular to the scattering plane and cannot be measured in the present experimental configuration. 
Precise X-ray measurements under high pressure 
in an extended $T$-range are now in progress.

The remarkable contrast between {\mord} and {\taf} below {\pcr} 
offers a test to the various theoretical scenarios for the 17.5 K transition. 
Let us first examine the possibility of a single transition 
at {\taf} ($=$ {\tord}) due to magnetic \hfill dipoles. 
\noindent
In general, {\taf} is derived from exchange interactions, 
and is independent of $g$. 
Therefore, the weak variations of {\taf} with pressure will be compatible with the ten-times increase of {\mord} $(= g\mu_{\rm B}m_{\rm o})$, 
only if $g$ is sensitive to pressure. 
The existing theories along this line explain the reduction of $g$ 
by assuming crystalline-electric-field (CEF) effects with 
low-lying singlets\cite{ge87}, and further by combining 
such with quantum spin fluctuations\cite{sikkema96,okuno98}.
To account for the $P$-increase of {\mord}, the characteristic energies 
of these effects should be reduced under pressure. 
Previous macroscopic studies however suggest 
opposite tendencies: the resistivity maximum shifts to higher temperatures 
\cite{mcelfresh87,ido93,schmidt93,oomi94} and
 the low-$T$ susceptibility decreases as $P$ increases\cite{louis86}.
The simple application of those models is thus unlikely to explain 
the behavior of {\mord} with pressure.

The models that predict a hidden (primary) non-dipolar order parameter 
$\psi$ are divided into two branches 
according to whether $\psi$ is odd (A) or 
even (B) under time reversal\cite{shah99}. 
The polarized neutron scattering has confirmed that the reflections arise 
purely from magnetic dipoles\cite{walker93}. 
For each branch, therefore, secondary order has been proposed as 
a possible solution of the tiny \hfill mo-
\noindent
ment. 
The Landau free energy for type (A) is given as 
\begin{equation}
F^{\rm (A)} = -\alpha(T_{\rm o} - T)\psi^{2} 
+ \beta\psi^{4} + Am^{2} - {\eta}m\psi, 
\end{equation}
where $\alpha, \beta$ and $A$ are positive, and the dimensionless order parameters $m$ and $\psi$ vary 
in the range $0 \leq m, \psi \leq 1$\cite{shah99}.
Minimization of $F^{\rm (A)}$ with respect to $m$ gives 
$m = - \delta\psi$, where $\delta \equiv \eta/2A$. 
The stability condition for $\psi$ then yields 
$\psi^{2} \sim \frac{\alpha}{2\beta}(T_{\rm o}' - T)$, 
where $T_{\rm o}' \sim T_{\rm o}(1 + {\cal O}(\delta^{2}))$.
If $\mu_{\rm para} \sim$ 1.2 {\mbohr} seen above 
{\tord}\cite{mason95} corresponds to 
$m \sim 1$, then the observed increase in {\mord} gives 
${\rm d}T_{\rm o}/{\rm d}P \sim T_{\rm o}{\rm d}m_{\rm o}^{2}/{\rm d}P 
\sim 0.8$ K/GPa, which is in 
good agreement with the experimental results 
 ($\sim 1.3$
K/GPa)\cite{louis86,mcelfresh87,fisher90,ido93,schmidt93,oomi94}.

In type (B), the simplest free energy invariant under 
time-reversal\cite{shah99,walker95} must take the form 
\begin{eqnarray}
F^{\rm (B)} =&& -\alpha(T_{\rm o} - T)\psi^{2} + \beta\psi^{4}\nonumber\\
&&+ a(T_{\rm m} - T)m^{2} + bm^{4} - {\zeta}m^{2}\psi^{2}. 
\end{eqnarray}
The continuous secondary order does not 
affect {\tord}, but enhances $C(T)$ at {\taf} 
as ${\Delta}C/T_{\rm m} \sim 
Nk_{\rm B}m_{\rm o}^{2}/T_{\rm m}$.
In the same way as in type (A), we obtain 
${\rm d}({\Delta}C/T_{\rm m})/{\rm d}P \sim$ 20 
mJ/K$^{2}$molGPa, 
when {\taf} $\sim$ {\tord}.
This cancels out with the $P$-increase in {\tord}, resulting in a roughly 
$P$-independent jump in $C(T)$.
This is consistent with previous $C(T)$ studies up to 0.6 
GPa\cite{fisher90}, in which ${\Delta}C_{\rm m}/T_{\rm o}$ is 
nearly constant, if entropy balance is considered.
Note that in type (B) {\taf} can in principle differ from {\tord}, 
which could also be consistent with the annealing effects\cite{fak96}. 

The phase transition at {\pcr} might be understood as a switching between $\psi$ and $m$ in type (B).
For example, the models of quadrupolar order in the CEF 
singlets $(\Gamma_{3}, \Gamma_{4})$\cite{santini94} and 
a non-Kramers doublet $(\Gamma_{5})$\cite{ami94,ohkawa99} both involve 
such magnetic instabilities. 
Interestingly, if the dipolar order takes place in the $\Gamma_{5}$ 
state, it will be accompanied by disappearance of magnon 
excitations, since the nature of excitations changes from dipolar origin to quadrupolar one\cite{ohkawa99}. 
Our preliminary results of inelastic neutron scattering support this 
possibility\cite{sato99}.

In conclusion, we have shown that the staggered magnetic moment 
associated with the 17.5 K transition in {\urs} is significantly 
enhanced by pressure. 
In contrast to the ten-times increase of the dipole moment, 
the transition temperature is insensitive to pressure. 
This feature is consistent with the hidden-order hypotheses. 
We have also found that the system undergoes a pressure-induced 
phase transition at around 1.5 GPa, evolving into a well-behaved 
magnetic phase.

We are grateful to T.\,Osakabe, T.\,Honma and Y.\,\={O}nuki for 
technical supports. 
One of us (H.A.) also thanks F.J.\,Ohkawa for helpful discussions. 
This work was partly supported by the JAERI-JRR3M Collaborative 
Research Program, and by Grant-in-Aid for Scientific Research from Ministry of Education, Science, Sports and Culture of Japan.

%
%

%
\end{document}